\documentclass[11pt]{article}
\usepackage{latexsym}
\usepackage{amsfonts,amssymb,amsmath}
\usepackage{txfonts}
\newtheorem{thm}{Theorem}[section]
\newtheorem{cor}[thm]{Corollary}
\newtheorem{lem}[thm]{Lemma}
\newtheorem{pro}[thm]{Proposition}
\newtheorem{defn}[thm]{Definition}

\title{ On the circuit-size of inverses }

\author{ J.C. Birget  }

\date{\today}
\begin{document}
\maketitle

\begin{abstract}

We reprove a result of Boppana and Lagarias: If 
$\Pi_2^{\sf P} \neq \Sigma_2^{\sf P}$ then there exists a partial
function $f$ that is computable by a polynomial-size family of 
circuits, but no inverse of $f$ is computable by a polynomial-size 
family of circuits.  We strengthen this result by showing 
that there exist length-preserving total functions that are one-way 
by circuit size and that are computable in {\em uniform} polynomial 
time.
We also prove, if $\Pi_2^{\sf P} \neq \Sigma_2^{\sf P}$, that
there exist polynomially balanced total {\em surjective} functions 
that are one-way by circuit size; here non-uniformity is used. 

\end{abstract}



\section{Introduction}

The difficulty of inversion (i.e., given $f$ and $y$, find any $x$ such 
that $f(x) = y$) is a fundamental topic in computational complexity 
and in cryptography. 
The question whether {\sf NP} is different from {\sf P} can be 
formulated as a question about the difficulty of inversion, namely, 
${\sf P} \neq {\sf NP}$ iff there exists a one-way function based on 
polynomial-time (\cite{Levin}, \cite{HemOgiCompan} pp.\ 32-43, 
\cite{DuKo} pp.\ 119-125). A function $f$ is said to be one-way based on
polynomial time iff $f$ is polynomial-time computable (by a deterministic
Turing machine) but no inverse function $f'$ of $f$ is polynomial-time 
computable. An {\it inverse} of $f$ is any function $f'$ that
$f \circ f' \circ f = f$.  
In this paper we consider one-way functions based on 
(non-uniform) families of circuits of polynomial size. 
Boppana and Lagarias \cite{BoppanaLagarias} (by using the Karp-Lipton
theorem \cite{KarpLipton}) proved that if 
$\Pi_2^{\sf P} \neq \Sigma_2^{\sf P}$ then there exists a partial function 
$f$ that can be computed by a non-uniform family of circuits of 
polynomial size, but no inverse $f'$ of $f$ can be computed by a 
non-uniform family of circuits of polynomial size. 
We show that this result still holds when $f$ is a total surjective and
polynomially balanced function, or when $f$ is length-preserving and 
uniformly computable in polynomial time (but non-uniformity is allowed 
for the inverses). 

\smallskip

By ``circuit'' we mean a digital circuit made of boolean gates, 
whose underlying directed graph is acyclic \cite{Wegener}.   
More precisely, a circuit $C$ with $m$ input vertices and $n$ output
vertices, consists of two parts. First, $C$ has an acyclic directed graph 
(with vertex set $V$ and edge set $E$); we assume that the set of vertices 
$V$ has a total order (i.e., $V$ is not just a set but a sequence).
Second, $C$ has a {\em gate map} 

\smallskip

 \ \ \ \ \ ${\sf gate}: \ \ v \in V \ \ \longmapsto \ \ {\sf gate}(v) \ $
$ \in \ $
$\{ {\sf and}, \ {\sf or}, \ {\sf not}, \ {\sf fork}, \ {\sf in}_1,$
$ \ \ldots, \ {\sf in}_m, \ {\sf out}_1, \ \ldots, \ {\sf out}_n \}$

\smallskip

\noindent which assigns a gate ${\sf gate}(v)$ to each vertex $v$. 
The gates {\sf and}, {\sf or}, and {\sf not} are the traditional boolean 
operations. The gates {\sf and} and {\sf or} have domain 
$\{0,1\} \times \{0,1\}$, so a vertex labeled by such a gate has 
in-degree 2; {\sf not} has domain $\{0,1\}$, so a vertex labeled by 
{\sf not} has in-degree 1; all three 
operations have codomain $\{0,1\}$, so the vertex has out-degree 1.
The gate 
 \ ${\sf fork}: x \in \{0,1\} \mapsto (x,x) \in \{0,1\} \times \{0,1\}$ 
 \ is also called the fan-out operation; the corresponding vertex has
in-degree 1 and out-degree 2.  Input vertices are mapped to 
${\sf in}_1, \ \ldots , {\sf in}_m$; they have in-degree 0 and 
out-degree 1. Output vertices are mapped to 
${\sf out}_1, \ \ldots , {\sf out}_n$; they have in-degree 1 and 
out-degree 0. The gate map is injective on the union of the set of input 
vertices and the set of output vertices.

The {\em size} (or complexity) of a circuit $C$, denoted $|C|$, is defined 
to be the number edges (i.e., wire links) plus the number of vertices. 
Thus $|C|$ is always at least as large as the number of input vertices, 
plus the number of output vertices.
A circuit $C$ with $m$ input vertices and $n$ output vertices has an 
{\it input-output function} 
 \ $(x_1, \ldots, x_m) \in \{0,1\}^m \ \longmapsto $
$ \ (y_1, \ldots, y_n) \in \{0,1\}^n$ that we denote by $C(.)$. 
The image set of $C$, i.e.\ the set all actual outputs, is denoted by 
${\sf im}(C)$ ($\subseteq \{0,1\}^n$).  

\smallskip

Let $A$ be a finite alphabet; when we talk about circuits we always
assume that $A = \{0,1\}$.

\begin{defn} \label{lep} \  A function $f: A^* \to A^*$ is called
{\em length-equality preserving} iff for all $x_1, x_2 \in A^*$,
 \ $|x_1| = |x_2|$ implies $|f(x_1)| = |f(x_2)|$. \ Equivalently,
for every $m$ there exists $n$ such that $f(A^m) \subseteq A^n$.

A special case consists of the {\em length-preserving} functions, 
satisfying \ $|f(x)| = |x|$. 
\end{defn}

\begin{defn} \label{polyBalanced} \ A function $f: A^* \to A^*$ is
called {\em polynomially balanced} iff there exist polynomials
$p_1(.)$ and $p_2(.)$  such that for all inputs $x \in A^*:$
 \ $|f(x)| \leq p_1(|x|)$ \ and \ $|x| \leq p_2(|f(x)|)$.

A special case is, again, the length-preserving functions.
\end{defn}

\begin{defn} \label{polyCircComp} \
A length-equality preserving function $f: \{0,1\}^* \to \{0,1\}^*$ 
is said to be {\em computed by a family of circuits}
${\bf C} = \{ C_m : m \in \varmathbb{N}\}$ iff for all
$m \in \varmathbb{N}$ and all $x \in \{0,1\}^m$, \ $f(x) = C(x)$.
(We do not make any uniformity assumptions for {\bf C}.)

This family is said to be of {\em polynomial size} iff 
there is a polynomial $p(.)$ such that for all $m:$ 
 \ $|C_m| \leq p(m)$. 
\end{defn}
In general, a family of circuits
${\bf C} = \{ C_i : i \in \varmathbb{N}\}$ could contain any 
number of circuits $C_i$ with the same number of input vertices; then
{\bf C} does not compute a function.

Computational {\it one-wayness} can be defined in many (non-equivalent) 
ways. We will use the following definition, related to worst-case circuit 
complexity (we are not considering cryptographic one-way functions 
here).  

\begin{defn} \label{onewayFuncSize} \  
A length-equality preserving function $f: \{0,1\}^* \to \{0,1\}^*$ is 
{\em one-way by circuit size} \ iff

\noindent $\bullet$ \ $f$ is polynomially balanced,

\noindent $\bullet$ \ $f$ is computable by a polynomial-size family of
circuits, but 

\noindent $\bullet$ \ no inverse function $f'$ of $f$ is computable by 
a polynomial-size family of circuits.
\end{defn}
Intuitively, one-wayness based on circuit size should be stronger than
one-wayness based on uniform computational complexity. Indeed, in
the former, not only is it difficult to find any inverse $f'$ of $f$,
but the circuits for the inverses $f'$ are all very large.
Definition \ref{onewayFuncSize} can also be adapted to a family of 
circuits, by itself. 

\begin{defn} \label{onewayCircFamily} \
A family of circuits ${\bf C} = \{ C_i : i \in \varmathbb{N}\}$
is {\em one-way by circuit size} \ iff
 \ for every polynomial $p(.)$ there is {\em no} family of circuits
${\bf C'} = \{ C_i' : i \in \varmathbb{N}\}$ such that
for all $i$, \  $C_i \circ C_i' \circ C_i(.) = C_i(.)$ \, and
 \, $|C'_i| \leq p(|C_i|)$.
\end{defn}

Before dealing with one-wayness we characterize the complexity of
the injectiveness problem and of the surjectiveness problem for 
circuits. 
Injectiveness is equivalent to the existence of left inverses, and
surjectiveness is equivalent to the existence of right inverses.
After that we consider general inverses.


\section{Injectiveness and surjectiveness}

The {\em equivalence problem} for circuits takes two circuits 
$C_1$, $C_2$ as input, and asks whether $C_1(.) = C_2(.)$. It is well 
known that this problem is {\sf coNP}-complete \cite{DuKo, HemOgiCompan}.
A related problem is the following, where for any set $S$ we 
denote the identity function on $S$ by ${\sf id}_S$.
In the {\it identity problem}, for a given circuit $C$ the question 
is whether $C(.) = {\sf id}_{ \{0,1\}^n }$.
In the {\it injectiveness problem} the question is whether $C(.)$ is 
injective. 
The identity problem is a special case of both the equivalence problem
and the injectiveness problem.

\begin{pro} \label{injectivenessProbl} \
The injectiveness problem and the identity problem for 
circuits are {\sf coNP}-complete.
\end{pro}
{\bf Proof} (this is Theorem 6.5 in \cite{BiRL}, reproved here
purely in the context of circuits).
 \ It is easy to see that the injectiveness problem and the identity 
problem are in {\sf coNP}. To show hardness we reduce the tautology 
problem for boolean formulas to the injectiveness problem and to the 
identity problem for circuits, as follows. 
Let $B$ be any boolean formula with $n$ variables. We define a new
boolean function $F_B: \{0,1\}^{n+1} \to \{0,1\}^{n+1}$ by

\smallskip

$F_B(x_1, \ldots, x_n, x_{n+1}) \ = \ $
$ \left\{ \begin{array}{ll}
(x_1, \ldots, x_n, x_{n+1}) \ \ \ \ \ \ \ &
  \mbox{if \ $B(x_1, \ldots, x_n) = 1$ \ or \ $x_{n+1} = 1$} , \\
  ( \, 1, \ \ldots, \ 1, \ 1) \ \ (= 1^{n+1}) & \mbox{otherwise.} 
  \end{array} \right.  $

\smallskip

\noindent Let us check that the following three properties are 
equivalent: 
 \ (1) \ $B$ is a tautology, \ (2) \ $F_B$ is injective, and 
 \ (3) \ $F_B = {\sf id}_{ \{0,1\}^{n+1} }$.
 
When $B(x_1, \ldots, x_n) = 1$ then
$F_B(x_1, \ldots, x_n, x_{n+1}) =$ $(x_1, \ldots, x_n, x_{n+1})$.  
So, if $B$ is a tautology then $F_B$ is the identity function on 
$\{0,1\}^{n+1}$ (which also implies that $F_B$ is injective).

If $B$ is a not a tautology then $B(c_1, \ldots, c_n) = 0$ for some
$(c_1, \ldots, c_n) \in \{0,1\}^n$. It follows that
$F_B(c_1, \ldots, c_n,0)$ $=$ $(1,\ldots, 1,1)$. But we also have
$F_B(1,\ldots, 1, 1) =$ $(1,\ldots,1,1)$, since here $x_{n+1} = 1$. 
Hence, $F_B$ is not injective (and hence not the identity function).
 \ \ \ \ \ $\Box$.

\bigskip

The {\em surjectiveness problem} for circuits takes a
circuit $C$ as input, and asks whether $C(.)$ is surjective.
Let $\Pi_2^{\sf P}$ denote the $\forall \exists$-class at level 2 in the 
polynomial hierarchy \cite{DuKo, HemOgiCompan}; similarly, 
$\Sigma_2^{\sf P}$ denotes the $\exists \forall$-class.
Theorem \ref{surjPi2} below is very similar to Theorem 5.9 in \cite{BiRL} 
about the surjectiveness problem for elements of the Thompson-Higman monoid 
$M_{2,1}$. But there are technical differences between circuits and 
elements of $M_{2,1}$, so we give a separate proof for circuits here.

\begin{thm} \label{surjPi2}
 \ The surjectiveness problem for circuits is $\Pi_2^{\sf P}$-complete.
\end{thm}
{\bf Proof.} \ The definition of surjectiveness shows that the surjectiveness
problem is in $\Pi_2^{\sf P}$. Indeed, $C(.)$ is surjective iff
 \ $(\forall y \in \{0,1\}^n)(\exists x \in \{0,1\}^m) \, [C(x) = y]$.
This is a $\Pi_2^{\sf P}$-formula, since $n, m \leq |C|$, and 
since the property $C(x) = y$ can be checked deterministically in 
polynomial time when $x$, $y$, and $C$ are given.

\smallskip

Let us prove hardness by reducing $\forall \exists{\sf Sat}$ (the 
$\forall \exists$-satisfiability problem) to the surjectiveness problem for
circuits. 
Let $B(x,y)$ be any boolean formula where $x$ is a sequence of $m$ boolean 
variables, and $y$ is a sequence of $n$ boolean variables. 
The problem $\forall \exists{\sf Sat}$ asks on input 
$\forall y \exists x \, B(x,y)$  whether this sentence is true. It is 
well known that $\forall \exists{\sf Sat}$ is $\Pi_2^{\sf P}$-complete 
\cite{DuKo, HemOgiCompan}.
We map the formula $B$ to the circuit $C_B$ with input-output 
function defined by

\smallskip

 \ \ \ $C_B(x, y, y_{n+1}) \ = \ \ $
  $\left\{ \begin{array}{ll}
      (y, y_{n+1})  \  & \mbox{if \ \ $B(x, y) = 1$ \ or \ $y_{n+1} = 1$,} \\
      (1^n,1) \ & \mbox{if \ \ $B(x, y) = y_{n+1} = 0$.}
           \end{array} \right. $

\smallskip

\noindent Equivalently, 

\smallskip

 \ \ \ $C_B(x, y, y_{n+1}) \ = \ $
$\big(y_1 \vee \overline{(B(x,y) \vee y_{n+1})}, \ \ \ldots \ , \ $
$y_n \vee \overline{(B(x,y) \vee y_{n+1})}, \ $
$y_{n+1} \vee \overline{B(x,y)} \, \big)$ . 

\smallskip

\noindent Hence one can easily construct a circuit for $C_B$ 
from the formula $B(x, y)$. By the definition of $C_B$ , 

\smallskip

 \ \ \ ${\sf im}(C_B) \ = \ \{(y,0) : \exists x B(x,y)\}$  $ \ \cup \ $
     $ \{(y,1) : y \in \{0,1\}^n\}$ \ \ ($ \ \cup \ \{(1^n,1) \}$) .

\smallskip

\noindent Since \, $(1^n,1) \in \{(y,1): y \in \{0,1\}^n\}$, the term 
$\{(1^n,1)\}$ (which may or may not be present) is irrelevant. Hence, 

\smallskip

 \ \ \ ${\sf im}(C_B) \ = \ $
 $\{0,1\}^n 1 \ \ \cup \ \ \{ y \in \{0,1\}^n : \exists x B(x,y)\} \, 0$ . 

\smallskip

\noindent Therefore, $\forall y \exists x \, B(x,y)$ is true \ iff \
${\sf im}(C_B) = \{0,1\}^n 1 \cup \{0,1\}^n 0$, i.e., iff $C_B$ is 
surjective.     \ \ \ $\Box$

\bigskip

For a partial function $f: X \to Y$ it is a well-known fact that $f$ is 
surjective iff $f$ has a right inverse.
By definition, a partial function $g:Y \to X$ is called a 
{\em right inverse} of $f$ iff $f \circ g(.) = {\sf id}_Y$. 
For circuits we have:
 \ A circuit $C$ (with $m$ input wires and $n$ output wires) is 
surjective iff there exists a circuit $C'$ (with $n$ input wires and $m$ 
output wires) such that \ \ $C \circ C'(.) = {\sf id}_{\{0,1\}^n}$. 

\begin{thm} \label{sizeRightInv} \ 
If there exists a polynomial $p(.)$ such that every surjective 
circuit $C$ has a {\em right inverse} $C'$ of size 
$|C'| \leq p(|C|)$, then $\Pi_2^{\sf P} = \Sigma_2^{\sf P}$ . 
\end{thm}
{\bf Proof.} \ If such a polynomial $p(.)$ exists then the surjectiveness 
of $C$ is characterized by 

\smallskip

 \ \ \ $C$ is surjective \ \  iff 
    \ \ $(\exists C', \, |C'| \leq p(|C|)) \, (\forall x \in \{0,1\}^m) \, $
  $[ C \circ C'(x) = x ]$.

\smallskip

\noindent This is a $\Sigma_2^{\sf P}$-formula since the quantified 
variables are polynomially bounded in terms of $|C|$, and the relation 
$C \circ C'(x) = x$ can be checked deterministically in polynomial time when 
$C$, $C'$ and $x$ are given.
This implies that the surjectiveness problem is in $\Sigma_2^{\sf P}$. But
since we already proved that the surjectiveness problem is 
$\Pi_2^{\sf P}$-complete, this implies that 
$\Pi_2^{\sf P} \subseteq  \Sigma_2^{\sf P}$.  
Hence, $\Pi_2^{\sf P} = \Sigma_2^{\sf P}$.
 \ \ \ $\Box$


\section{General inverses}

The general concept of an inverse goes back to Moore \cite{Moore} 
(Moore-Penrose pseudo-inverse of a matrix), and von Neumann \cite{vNeu} 
(regular rings). 
For a partial function $f: X \to Y$, the domain of $f$
is denoted by ${\sf dom}(f)$ \, ($\subseteq X$), and the image (or range)
is denoted by ${\sf im}(f)$ \, ($\subseteq Y$). A partial function
$f: X \to Y$ is called {\em total} iff ${\sf dom}(f) = X$.
When we just say ``function'' we mean a total function.

\begin{defn} \label{defInverse} \      
For a partial function $F: X \to Y$ an {\em inverse} (also called a
{\em semi-inverse}) of $F$ is any partial function $F': Y \to X$ such 
that $F \circ F' \circ F = F$.
If both $F \circ F' \circ F = F$ and $F' \circ F \circ F' = F'$ hold then
$F'$ is a {\em mutual inverse} of $F$, and $F$ is a mutual inverse of $F'$.
\end{defn}
The following facts about inverses are well known and 
straightforward to prove. 
For any two partial functions $F: X \to Y$ and $F': Y \to X$ we have: 

\smallskip

{\em  
\noindent $\bullet$ \ \ $F \circ F' \circ F = F$ \ \ iff 
 \ \ $(F \circ F')_{ {\sf im}(F)} \, = \, {\sf id}_{ {\sf im}(F)}$ ,
 \ where $(.)_{ {\sf im}(F)}$ denotes the restriction to ${\sf im}(F)$.

\smallskip

\noindent $\bullet$ \ \ If $F'$ is a semi-inverse of $F$ then
 \ ${\sf im}(F) \subseteq {\sf dom}(F')$; 
 \, i.e., $F'(y)$ is defined for all $y \in {\sf im}(F)$.

\smallskip

\noindent $\bullet$ \ \ If $F'$ is a semi-inverse of $F$ then
 \ $F'_{ {\sf im}(F) }$ is injective. 

\smallskip

\noindent $\bullet$ \ \ If $F'$ is a semi-inverse of $F$ then 
$F' \circ F \circ F'$ is a mutual inverse of $F$.   
 
\smallskip

\noindent $\bullet$ \ \ Every partial function $F$ has at least one 
semi-inverse. More specifically, $F$ has at least one semi-inverse 
$F'_1$ that is total (i.e., ${\sf dom}(F'_1) = Y$), and at least one 
semi-inverse $F'_2$ that is injective and whose domain is ${\sf im}(F)$.
}  

For infinite sets the last fact requires the axiom of choice.
The following two Lemmas are also straightforward. 

\begin{lem} \label{RinvIFFmutinvtotinj} \ 
$F'$ is a right inverse of $F$ \ iff \ $F'$ is a total and injective 
 mutual inverse of $F$. \ \ \ \ \ $\Box$
\end{lem}

\begin{lem} \label{CharacterizSurj2} \    
For a partial function $F: X \to Y$ the following are equivalent:

\smallskip

\noindent {\bf (1)} \ \ $F$ is {\em surjective};

\smallskip

\noindent {\bf (2)} \ \ $F$ has a right inverse;

\smallskip

\noindent {\bf (3)} \ \ $F$ has a mutual inverse $F'$ that is total 
and injective; 

\smallskip

\noindent {\bf (4)} \ \ every semi-inverse $F'$ of $F$ is total
and injective;

\smallskip

\noindent {\bf (5)} \ \ every semi-inverse $F'$ of $F$ is total.
 \ \ \ \ \ $\Box$ 
\end{lem}
We can now reformulate Theorem \ref{sizeRightInv} in terms of inverses.

\begin{thm} \label{sizeInverses} \    
If there exists a polynomial $p(.)$ such that every circuit
$C$ has a semi-inverse $C'$ of size $|C'| \leq p(|C|)$, then
$\Pi_2^{\sf P} = \Sigma_2^{\sf P}$ .
\end{thm}
{\bf Proof.} \ If such a $p(.)$ exists then every circuit $C$ has an 
inverse $C'$ of size $|C'| \leq p(|C|)$, and hence every $C$ has a mutual 
inverse $C'_2 = C' \circ C \circ C'$ of size 
$|C_2'|  \leq 2 \cdot p(|C|) + |C|$. 
Let $q(n) = 2 \cdot p(n) + n$, which is also a polynomial. 

Let us now consider the special case where $C$ is surjective. Then by
Lemma \ref{CharacterizSurj2}(1 $\Rightarrow$ 4), $C'_2$ is total and 
injective. Then by Lemma \ref{RinvIFFmutinvtotinj}, since $C'_2$ is a mutual 
inverse, $C'_2$ is a right inverse of $C$. 
Now Theorem \ref{sizeRightInv} (for the polynomial $q(.)$) implies that 
$\Pi_2^{\sf P} = \Sigma_2^{\sf P}$. 
  \ \ \ $\Box$

\medskip 

\noindent Theorem \ref{sizeInverses} is not new; it follows 
immediately from a result by Boppana and Lagarias (Theorem 2.1a in
\cite{BoppanaLagarias}), combined with the Karp-Lipton Theorem 
\cite{KarpLipton, DuKo, HemOgiCompan}. 

The proof of Theorem \ref{sizeInverses} also applies to surjective 
functions (while the methods in \cite{BoppanaLagarias} do not seem
to): 

\begin{cor} \label{sizeInversesSurj} \    
If there exists a polynomial $p(.)$ such that every {\em surjective} 
circuit $C$ has a {\em semi-inverse} $C'$ of size \ $|C'| \leq p(|C|)$, 
then $\Pi_2^{\sf P} = \Sigma_2^{\sf P}$ .  \ \  \ \ \ $\Box$ 
\end{cor} 
Theorems \ref{sizeRightInv}, \ref{sizeInverses} and Coroll.\
\ref{sizeInversesSurj} show that the family of all circuits and the 
family of all surjective circuits are one-way by circuit-size.


\section{One-way functions, if  $\Pi_2^{\sf P} \neq \Sigma_2^{\sf P}$}

We will use the above results to construct two types of functions that 
are one-way by circuit-size.

\subsection{A surjective non-uniform one-way function}

The papers \cite{FFNR} and \cite{BFKRV} discuss the existence of 
surjective one-way functions, based on uniform polynomial time 
complexity. In the uniform case (with uniformity for both $f$ and
$f'$), it is known that ${\sf P} \neq {\sf NP} \cap {\sf coNP}$ 
implies the existence of one-way functions (attributed to 
\cite{BorodinDemers} in the Introduction of \cite{FFNR}).
Here we give an existence result for surjective one-way functions 
with respect to non-uniform polynomial time, i.e., circuit size.  

For a circuit $C$ we will denote the number of input vertices by $m_C$ 
or $m$, and the number of output wires by $n_C$ or $n$.  
An {\it identity wire} in a circuit is an edge $(x_i, y_j)$ that 
directly connects an input vertex $x_i$ to an output vertex $y_j$; so 
$x_i$ and $y_j$ have the same value. To add an identity wire means to 
create a new input vertex, a new output vertex, and an edge between
them.

\begin{lem} \label{idwiresSurj} \   
Suppose $C_0$ is obtained from $C$ by adding identity wires. Then
$C_0$ is surjective iff $C$ is surjective.
\end{lem}
{\bf Proof.} \ Let $j$ be the number of identity wires added.
So, ${\sf im}(C_0) = {\sf im}(C) \times \{0,1\}^j$. 
Then $C$ is surjective iff \ ${\sf im}(C) = \{0,1\}^n$ \ iff 
 \ ${\sf im}(C_0) = \{0,1\}^n \times \{0,1\}^j = \{0,1\}^{n+j}$
 \ iff $C_0$ is surjective.   \ \ \ $\Box$

\begin{pro} \label{ClinsizeinNsurj} \ 
Theorem \ref{sizeRightInv} and Corollary \ref{sizeInversesSurj} 
still hold when one only considers surjective circuits $C$ that 
satisfy \, $m \leq \frac{1}{2} \, |C| < 2n$.
 \ The same holds if one considers only surjective circuits that 
satisfy \, $2 n < m \leq |C| < 6n$.
\end{pro}
{\bf Proof.} \ From any circuit $C$ one can construct a circuit 
$C_1$ by adding $|C|$ identity wires. Then $C$ is surjective iff 
$C_1$ is surjective (by Lemma \ref{idwiresSurj}).
An identity wire has two vertices and one edge, so the resulting
circuit $C_1$ has size $|C_1| = 4 \, |C|$. For the number of input 
vertices and output vertices we have \ $m_1 = m + |C|$, 
and \ $n_1 = n + |C|$. Since $m \leq |C|$, it follows that 
$m_1 \leq \frac{1}{2} \, |C_1|$. 
Also, \ $|C_1| = 4 \, (n_1 - n) < 4 \, n_1$.  

The circuit $C_1$ satisfies $2n_1 > m_1$ \, (since 
$m_1 \leq  \frac{1}{2} \, |C_1| < 2n_1$). 
Now $2n_1 - m_1 +1$ new input vertices can be added to $C_1$; 
these vertices are not connected to anything and are not output
vertices. Then the new circuit $C_2$ is surjective iff $C_1$ is 
surjective. The new circuit $C_2$ satisfies \ $n_2 = n_1$, 
 \ $|C_2| = |C_1| + 2n_1 - m_1 + 1 \leq |C_1| + 2n_1$
$ < 4n_1 + 2n_1$, and \ $m_2 = 2n_1 + 1 > 2 \, n_2$. Hence,
$2 \, n_2 < m_2 \leq |C_2| \leq 6 \, n_2$.  

The circuits $C_1$ and $C_2$ can be constructed from $C$ 
deterministically in polynomial time.  Moreover, an inverse of 
$C$ can be obtained in polynomial time from
an inverse of $C_1$, and vice versa. The same holds for $C_2$.
Hence, $C$ has an inverse of size $\leq p(|C|)$ \, (for some 
polynomial $p(.)$) \, iff \, $C_i$ has an inverse of size 
$\leq p_i(|C_i|)$ (for some polynomial $p_i(.)$, $i = 1, 2$).
Since the existence of polynomial-size inverses for all surjective
circuits $C$ implies $\Pi_2^{\sf P} = \Sigma_2^{\sf P}$ (by Corollary
\ref{sizeInversesSurj}), the existence of polynomial-size inverses 
for $C_1$ or $C_2$ also implies $\Pi_2^{\sf P} = \Sigma_2^{\sf P}$.
 \ \ \ \ \ $\Box$

\medskip

We saw in Lemma \ref{CharacterizSurj2} that a function 
$f: X \to Y$ is surjective iff every inverse of $f$ is total and 
injective.

\begin{thm} \label{sizeRightInvInfOften} \
For every polynomial $p(.)$ consider the following set of 
{\em surjective} circuits:

\smallskip

\hspace{0.6in} ${\bf C}_p \ = \ \big\{ \, C \ :$
  $ \ 2 \, n_C < m_C \leq |C| < 6 \, n_C$ 
  \ {\rm and every inverse} $C'$ {\rm of} $C$ {\rm satisfies}
  \ $|C'| > p(|C|) \, \big\}$.

\smallskip

\noindent
If $\Pi_2^{\sf P} \neq \Sigma_2^{\sf P}$ then for every polynomial
$p(.)$ the set
 \ $\{n_C : C \in {\bf C}_p\}$ \ (consisting of the output 
lengths of the circuits in ${\bf C}_p$)  is {\em infinite}. 
\end{thm}
{\bf Proof.} \ We assume $\Pi_2^{\sf P} \neq \Sigma_2^{\sf P}$.  
Then by Corollary \ref{sizeInversesSurj} and Prop.\ 
\ref{ClinsizeinNsurj}, ${\bf C}_p$ is not empty. For all 
$C \in {\bf C}_p$ we have $2 \, n_C < m_C \leq |C| < 6 \, n_C$. 
It follows that for any polynomial $p(.)$ the four sets 
 \ ${\bf C}_p$, \ $\{|C| : C \in {\bf C}_p\}$, 
 \ $\{n_C : C \in {\bf C}_p\}$, and \, $\{m_C : C \in {\bf C}_p\}$ 
 \, are all infinite iff one of them is infinite.
Moreover, if a function is surjective then all its inverses are total
and injective (Lemma \ref{CharacterizSurj2}).  Hence, ${\bf C}_p$ is 
infinite iff the set 
 \ $\{C': C'$ {\sl is an inverse of some} $C \in {\bf C}_p \}$ \ is
infinite, iff the set
 \ $\{|C'| : C'$ {\sl is an inverse of some} $C \in {\bf C}_p \}$ 
 \ is infinite.

For two polynomials we write $p_2 \geq p_1$ when $p_2(n) \geq p_1(n)$ 
for all $n$.  If $p_2 \geq p_1$ then
${\bf C}_{p_2} \subseteq {\bf C}_{p_1}$; hence for any polynomial
$p_0(.)$ we have 
 \ $\bigcup_{p \geq p_0} {\bf C}_p \ = \ {\bf C}_{p_0}$. 
For any polynomial $p_0(.)$ the set

\smallskip

\hspace{1in} {\sl $\{ p(|C|) : \ p(.)$ is a polynomial, \ $p \geq p_0$, 
                \ and \ $C \in {\bf C}_p\}$ }

\smallskip

\noindent is infinite; indeed, the set of polynomials is infinite and 
each ${\bf C}_p$ is non-empty.
It follows that for any polynomial $p_0(.)$ the set
 \ {\sl $\{|C'| : C'$ is an inverse of some $C \in {\bf C}_p$, for some 
$p \geq p_0\}$}  \ is infinite, since $|C'| > p(|C|)$ when 
$C \in {\bf C}_p$.
Hence, for any  $p_0(.)$, \,  ${\bf C}_{p_0}$ \, and 
 \, $\{ n_C : C \in {\bf C}_{p_0}\}$ \, are infinite.  
 \ \ \ \ \ $\Box$

\begin{thm} \label{surjOneW} \  If $\Pi_2^{\sf P} \neq \Sigma_2^{\sf P}$
then there exists a {\em surjective} total function $f: \{0,1\}^* \to
\{0,1\}^*$ which is polynomially balanced and length-equality preserving,
and which satisfies:

\smallskip

\noindent $\bullet$ \ \ $f$ is computed by a non-uniform
polynomial-size family of circuits, but 

\smallskip

\noindent $\bullet$ \ \ $f$ has no inverse that can be computed by a 
non-uniform polynomial-size family of circuits.  
\end{thm} 
{\bf Proof.} \ Consider an infinite sequence of polynomials
 \ $p_1 < p_2 < \ \ldots \ < p_k < \ \ldots $ \ , with 
$p_k(x) > x^k + k$ for all numbers $x$.  Recall that 
 \ ${\bf C}_{p_k} \ = \ \big\{ \, C \, :$
  $ \ 2 \, n_C < m_C \leq |C| < 6 \, n_C$
  {\sl and every inverse $C'$ of $C$ satisfies}
  \ $|C'| > p_k(|C|) \, \big\}$. 
 \ Let us abbreviate ${\bf C}_{p_k}$ by ${\bf C}_k$.
We saw that \ $\ \ldots \ \subseteq {\bf C}_k \subseteq \ \ldots \ $
$\subseteq {\bf C}_2 \subseteq {\bf C}_1$. 
By Theorem \ref{sizeRightInvInfOften}, if 
$\Pi_2^{\sf P} \neq \Sigma_2^{\sf P}$ then ${\bf C}_k$ and 
$\{|C| : C \in {\bf C}_k\}$ are infinite for every $k$.  
We now construct an infinite set of circuits
$\{C_k \in {\bf C}_k: k \in {\bf N}\}$, where we abbreviate
$m_{C_{k}}$ and $n_{C_{k}}$ by $m_k$, respectively $n_k$.

\smallskip

$C_1$ is a smallest circuit in ${\bf C}_1$ ;

\smallskip

$C_{k+1}$ is a smallest circuit in 
 \ \  $\{ C \in {\bf C}_{k+1} : \ |C| > |C_k| , $
   $ \ \ n_C > 1 + n_k \ \ {\rm and} $
   $ \ m_C > 2 \, m_k \}$ .

\smallskip

\noindent Since ${\bf C}_{k+1}$ is infinite (by Theorem
\ref{sizeRightInvInfOften}), the circuit $C_{k+1}$ exists. 

\smallskip

\noindent {\sf Claim:} 
 \ \ \ $m_{k+1} - m_k \ > \ n_{k+1} - n_k \ > \ 1$.

\noindent {\sf Proof of the Claim:} 
 \ We have $m_{k+1} > 2m_k$ (by the choice of $C_{k+1}$), and 
$m_{k+1} > 2n_{k+1}$ (since $C_{k+1} \in {\bf C}_{k+1}$). 
Hence, $m_{k+1}/2 > m_k$ and $m_{k+1}/2 > n_{k+1}$. 
By adding these inequalities we obtain $m_{k+1} > m_k + n_{k+1}$, 
hence $m_{k+1} - m_k > n_{k+1} > n_{k+1} - n_k$. 
Also, the choice of $n_C > 1 + n_k$ implies $n_{k+1} - n_k > 1$.
This proves the Claim.

\medskip
 
\noindent We define a total and surjective function 
$F: \{0,1\}^* \to \{0,1\}^*$ as follows:

\smallskip

(1) \hspace{.6in} 
                $F(x) \ = \ C_k(x)$ \ \ if \ $|x| = m_k$; 

\medskip

(2) \hspace{.6in} $F$ \, maps \, 
$D_k = \bigcup_{m \, = \, m_k +1}^{m_{k+1}-1} \{0,1\}^m$ \, onto
 \, $R_k = \bigcup_{n \, = \, n_k +1}^{n_{k+1}-1} \{0,1\}^n$.

\medskip

\noindent In (1), $F$ maps $\{0,1\}^{m_k}$ onto $\{0,1\}^{n_k}$ for 
every $k$, since $C_k$ is surjective.
In (2), $D_k$ and $R_k$ are non-empty, since 
$m_{k+1} - m_k > n_{k+1} - n_k \geq 1$ (by the Claim).
To complete the definition of $F$, $D_k$ can be mapped onto $R_k$ 
in a length-equality preserving way, as follows:  
Since $m_{k+1} - m_k -1 > n_{k+1} - n_k -1$ and $m_i > n_i$ 
(for all $i$), we can map $\{0,1\}^{m_k + i}$ onto 
$\{0,1\}^{n_k + i}$ for
$i = 1, \ \ldots, \ n_{k+1} - n_k -1$ ($\leq m_{k+1} - m_k -1$).
Next, we map \ $\bigcup_{n_k \leq m < m_{k+1}} \{0,1\}^m$ \, onto 
$\{0,1\}^{n_{k+1} -1}$. This way, $F$ is onto and length-equality 
preserving. In more detail yet, when $j > i$ we map $\{0,1\}^j$ 
onto $\{0,1\}^i$ by 
 \ $(x_1, \ldots, x_i, x_{i+1}, \ldots, x_j) \mapsto $
$(x_1, \ldots, x_i)$.
This way, $F: D_k \to R_k$ consists of projections.

Let us check that overall, $F$ is polynomially balanced (in fact,
input sizes and output sizes bound each other linearly): Indeed,
$F$ maps length $m_k$ to length $n_k$, with $m_k < 6n_k$. Also,
length $m_k + i$ is mapped to $n_k + i$ for 
$1 \leq i < n_{k+1} - n_k$, with $m_k + i < 6n_k + i$.  
Finally, lengths between $m_k + n_{k+1} - n_k$ and $m_{k+1} -1$ 
are mapped to length $n_{k+1} -1$, with 
$m_{k+1} -1 < 6 n_{k+1} -1$.

We see that $F$ can be computed by a linear-size non-uniform 
family of circuits:  For inputs of length $m_k$ (for some $k$) 
we use the circuits $C_k$; for the other inputs, $F$ is a 
projection.

Finally, let us check that no inverse $F'$ of $F$ is computable 
by a polynomial-size circuit family (if 
$\Pi_2^{\sf P} \neq \Sigma_2^{\sf P}$). 
The set $\{C_k : k \in \varmathbb{N}\}$ that we constructed is 
infinite and $C_k \in {\bf C}_k$; hence any family 
$(C'_k : k \in \varmathbb{N})$ of circuits that computes an inverse 
$F'$ will satisfy $|C'_k| > |C_k|^k + k$ for all $k$. 
Since the set $\{ n_k : k \in \varmathbb{N}\}$ is infinite, the 
restriction of $F$ to 
 \ $\bigcup_{k \in \varmathbb{N}} \{0,1\}^{m_k} \ \to \ $
$\bigcup_{k \in \varmathbb{N}} \{0,1\}^{n_k}$ \ has no inverse with
size bounded by a polynomial (of fixed degree). Thus $F$ has no 
polynomial-size inverse.  \ \ \ \ \ $\Box$

\subsection{A uniform one-way function}

A result of Boppana and Lagarias \cite{BoppanaLagarias} (combined 
with the Karp-Lipton theorem \cite{KarpLipton}) states that if 
$\Pi_2^{\sf P} \neq \Sigma_2^{\sf P}$ then there exists a function 
$f$ that is one-way in the sense that $f$ computable by a 
polynomial-size family of circuits, but the inverses of $f$ are not
computable by any polynomial-size family of circuits. The one-way 
functions considered in \cite{BoppanaLagarias} are not polynomially 
balanced; moreover, they are either not total or not length-equality 
preserving (in the terminology of \cite{BoppanaLagarias}, the output 
can be the single symbol $\#$). Also, these one-way functions are 
based on the Karp-Lipton theorem, so they are (apparently) not 
computable in uniform polynomial time. 
We will now construct a length-preserving function $f$ that can be 
computed uniformly in polynomial time, but no inverse $f'$ has a 
polynomial-size family of circuits. 

We can describe any circuit $C$ by a bitstring ${\sf code}(C)$, i.e.,  
there is a ``G\"odel numbering'' for circuits.
Naturally there is also a decoding function ${\sf decode}(.)$ which is
an inverse of ${\sf code}(.)$, i.e., ${\sf decode}({\sf code}(C)) = C$.
We can extend  ${\sf decode}(.)$ to a total function, so any bitstring 
is decoded to a circuit.
The encoding function ${\sf code}(.)$ is associated with an
{\em evaluation function} {\sf ev} such that

\smallskip

 \hspace{.4in} ${\sf ev}\big({\sf code}(C), \, x \big) \ = \ C(x)$ 
 \ \ for all $x \in \{0,1\}^{m_c}$.

\smallskip

\noindent Here we denote the length of the inputs of $C$ by $m_C$
and the length of the outputs by $n_C$. 
The functions ${\sf code}(.)$, ${\sf decode}(.)$, and ${\sf ev}(.,.)$ 
can be constructed so that they have special properties. 
The existence and the main properties of ${\sf ev}(.,.)$ and 
${\sf code}(.)$ are well-known folklore, but we prove them here 
nevertheless because we will need detailed size and complexity 
estimates (items 3, 4, and 5 in the Proposition below).

\begin{pro} \label{codeANDevProp} \ Let ${\bf C}$ denote the
set of all circuits.  There exist functions

\smallskip

\hspace{.5in} ${\sf code}: \ {\bf C} \ \to \ \{0,1\}^*$ ,

\smallskip

\hspace{.5in} ${\sf decode}: \ \{0,1\}^* \ \to \ {\bf C}$ ,

\smallskip

\hspace{.5in}
${\sf ev}: \ \{0,1\}^* \times \{0,1\}^* \ \to \ \{0,1\}^*$ , 
 \ \ \ such that

\medskip

\noindent {\rm (1)} \ \ \ \ for all $C \in {\bf C}:$
 \ \ \ ${\sf decode}({\sf code}(C)) = C$ ;

\smallskip

\noindent {\rm (2)} \ \ \ \ for all $c, x \in \{0,1\}^*$ with
 \ $|x| = m_{{\sf decode}(c)}:$
 \ \ \ \ ${\sf ev}(c, x) \ = \ [{\sf decode}(c)](x)$ ;

\smallskip

 \ \ \ \ in particular, for all \, $C \in {\bf C}$,  
 \, $x \in \{0,1\}^{m_C}:$
 \ \ \ \ ${\sf ev}\big({\sf code}(C), \, x \big) \ = \ C(x)$ ;

\smallskip

\noindent {\rm (3)} \ \ \ \ for all $C \in {\bf C}:$ 
 \ \ \ $|C| \, \log_2 |C| \ < \ $
$|{\sf code}(C)| \ $
$ < \ 6 \, |C| \, \log_2 |C|$ ;

%

\smallskip

\noindent {\rm (4)} \ \ \ \ ${\sf decode}(.)$ and ${\sf ev}(.,.)$ are 
total functions;

\smallskip

\noindent {\rm (5.1)} \ \ the language
 \ ${\sf im}({\sf code}) \ = \ {\sf im}({\sf code} \circ {\sf decode})$
 $ \ \subseteq \{0,1\}^*$ \ belongs to {\sf P};

\smallskip

\noindent {\rm (5.2)} 
 \ \ ${\sf code} \circ {\sf decode}(.): \{0,1\}^* \to \{0,1\}^*$
 \ is polynomial-time computable and polynomially balanced;

\smallskip

\noindent {\rm (5.3)} \ \ ${\sf ev}(.,.)$ \, is polynomial-time computable.
\end{pro}
{\bf Proof.} \ We denote the sets of vertices and edges of $C$ by $V$,
respectively $E$.  To construct the bitstring ${\sf code}(C)$ from a
circuit $C$ we first use a four-letter alphabet $\{a,b,c,d\}$.
We label the vertices of the acyclic digraph of $C$ injectively by 
strings over $\{a,b\}$, using binary numbering (with $a = 0$, $b = 1$),
according to the order of $V$, from number 0 through $|V| -1$.
Each vertex is thus represented by a string in $\{a,b\}^*$ of length 
$\lceil \log_2 |V| \rceil$.
In addition, each vertex is labeled by its gate type (namely {\sf and}, 
{\sf or}, {\sf not}, {\sf fork}, ${\sf in}_1, \ \ldots, {\sf in}_m$,
${\sf out}_1, \ \ldots, {\sf out}_n$), according to the gate map; 
strings over $\{c,d\}$ of length $\lceil \log_2 (4+m+n) \rceil$ 
are used for these gate labels. As we alternate between $\{a,b\}$ and 
$\{c,d\}$, no separator is needed. Thus we have a description of length 
 \ $|V| \, (\lceil \log_2 |V| \rceil + \lceil \log_2 (4+m+n) \rceil)$
 \ for the list of vertices and their gate types.
Each edge is described by a pair of vertex codes, separated 
by a letter $c$, and any two edges are separated by a letter $d$. 
Thus the list of edges is described by a string of length 
 \ $|E| \, (2 + 2 \, \lceil \log_2 |V| \rceil)$.
So, \ $|{\sf code}(C)| \ = $
 \ $|V| \, (\lceil \log_2 |V| \rceil + \lceil \log_2 (4+m+n) \rceil)$
 \ $ \ + \ |E| \, (2 + 2 \, \lceil \log_2 |V| \rceil)$. 
Hence \ $|{\sf code}(C)| \ > \ \frac{1}{2} \, |C| \, \log_2 |C|$ 
 \ (since $|V|^2 + |V| \geq |E| + |V| = |C|$), and 
 \ $|{\sf code}(C)| \ < \ 3 \, |C| \, \log_2 |C|$
 \ (since $|C| = |V| + |E|$).
Turning ${\sf code}(C)$ into a bitstring (e.g., by encoding $a, b, c, d$ 
as 00, 01, 10, 11, respectively) doubles the length. This completes the 
definition of ${\sf code}(.)$ and proves property (3).  

To define the function {\sf decode} we first let
${\sf decode}({\sf code}(C)) = C$.  When $c$ is not the code of any 
circuit, we let ${\sf decode}(c)$ be the largest identity circuit 
(i.e., computing the identity map on $\{0,1\}^m$, for some $m$) with 
a code of length $\leq |c|$.
This makes ${\sf decode}(.)$ a total function; property (1) also 
follows immediately. 

An evaluation function {\sf ev} can now be defined, based on the 
above construction of ${\sf code}(.)$ and ${\sf decode}(.)$.
For any $(c,x) \in \{0,1\}^* \times \{0,1\}^*$, let $C = {\sf decode}(c)$.  
If $|x| = m_C$ then we define 
 \ ${\sf ev}(c, x) \ = \ [{\sf decode}(c)](x)$. 
If $|x| \neq m_C$ we define \ ${\sf ev}(c, x) \ = \ x$.
Properties (2) and (4) now hold.

The definitions of {\sf code} and {\sf decode} make it easy to check
whether a string $c$ is an encoding of a circuit, and to decode $c$
(or to generate an identity circuit if $c$ is not a code).
The inequalities in (3) imply that ${\sf decode} \circ {\sf code}(.)$ 
is polynomially balanced. 
This shows properties (5.1) and (5.2). 
The definitions of {\sf code}, {\sf decode}, and {\sf ev} make it easy 
to compute ${\sf ev}(c,x)$, so we have (5.3). The details are very 
similar to the proof that the circuit value problem is in {\sf P} 
(see section 4.3 of \cite{Papadim}).
 \ \ \ $\Box$

\medskip

\noindent The function {\sf ev} is neither length-equality preserving 
nor polynomially balanced.

\begin{pro} \label{ClinsizeinN} \
Let $m$ and $n$ denote, respectively, the number of input and output
vertices of a circuit $C$.
Theorem \ref{sizeInverses} still holds when one only considers
circuits $C$ that satisfy \, $|C| < 2m$ \, and \, $m = n$ \,
(i.e., the function $C(.)$ is length-preserving).

Theorem \ref{sizeInverses} also holds when one only considers
circuits $C$ with \, $m = n$ \, and
 \ $|{\sf code}(C)| \ < \ 12 \, m \, \log_2 (2 \, m)$.
\end{pro}
{\bf Proof.} \ From $C$ one can construct a circuit $C_1$ with equal
numbers of input and output vertices. If $m < n$ one adds $n - m$
extra input vertices that are not connected to anything else in the
circuit.  If $m > n$ one adds $m - n$ new output vertices that carry
the constant boolean value 0. A constant 0 can be created by making
two copies of the input $x_1$ (by forking twice) and then taking
$x_1 \wedge \overline{x_1}$ (= 0); this uses 4 gates and 6 wires.
Making $m - n -1$ more copies of 0 uses $m - n -1$ {\sf fork} gates
and $2 \, (m - n -1)$ more wires.
Now $m_1 = n_1 = {\sf max}\{n, m \}$, and
$|C_1| \leq |C| + 3 \, |m - n| + 10$ \, (where $|m - n|$ denotes
the absolute value of $m - n$).
Inverting $C$ is equivalent to inverting $C_1$.

In any circuit $C_1$ one can add $|C_1|$ identity wires.
An identity wire has two vertices and one edge, so the resulting
circuit $C_2$ has size $|C_2| = 4 \, |C_1|$, and
 \ $m_2 = m_1 +  3 \, |C_1|$ input vertices, and
$n_2 = n_1 + 3 \, |C_1|$ output vertices. Hence, $|C_2| < m_2 + n_2$.
Recall that circuit size is defined to be the number of vertices plus
the number of edges in the circuit. If $m_1 = n_1$ then $m_2 = n_2$,
and $|C_2| < 2 \, m_2$.
Since $C_1$ and $C_2$ differ only by identity wires, there is a
one-to-one correspondence between inverses of $C_1$ and of $C_2$; an
inverse for $C_2$ can be obtained from an inverse of $C_1$ by adding
identity wires; an inverse for $C_1$ can be obtained from an inverse of
$C_2$ by removing the extra identity wires.

By Prop.\ \ref{codeANDevProp}, $C_2$ also satisfies
 \ $|{\sf code}(C_2)| \leq 6 \, |C_2| \, \log_2 |C_2|$. We saw that
$|C_2| < 2 \, m_2$, hence
 \ $|{\sf code}(C_2)| < 12 \, m_2 \, \log_2 (2 \, m_2)$.

The circuits $C_1$ and $C_2$ can be constructed from $C$
deterministically in polynomial time.
Moreover, an inverse of $C$ can be obtained in polynomial time from
an inverse of $C_1$ or $C_2$, and vice versa.  Hence, $C$ has an
inverse of size $\leq p(|C|)$ \, (for some polynomial $p(.)$)
 \, iff \, $C_i$ has an inverse of size $\leq p_i(|C|)$
(for some polynomial $p_i(.)$), $i = 1,2$.
Since the existence of polynomial-size inverses for all circuits $C$
implies $\Pi_2^{\sf P} = \Sigma_2^{\sf P}$ (by Theorem \ref{sizeInverses}),
the existence of polynomial-size inverses for circuits $C_i$ also implies
$\Pi_2^{\sf P} = \Sigma_2^{\sf P}$.
 \ \ \ $\Box$

\bigskip

Based on Propositions \ref{codeANDevProp} and \ref{ClinsizeinN} we 
now construct a function which is one-way by circuit size. We start 
with the function

\smallskip

\hspace{0.6in}          ${\sf ev}_{\sf circ}: \ \ \ $
$(c, \ x) \ \longmapsto \ \big(c, \ [{\sf decode}(c)](x) \, \big)$ 

\smallskip

\noindent which is just the pairing 
$\langle \pi_1, \, {\sf ev} \rangle$ of the first projection
 \ $\pi_1: (x_1,x_2) \longmapsto x_1$ \ and the evaluation 
function {\sf ev}.
We saw that {\sf ev} is a total function that can be computed 
deterministically in polynomial time, hence ${\sf ev}_{\sf circ}$ 
is also total and polynomial-time computable.
Levin observed that ${\sf ev}_{\sf circ}$ is a complete or 
``universal'' one-way function, for a certain definition of 
one-way functions and for certain reductions between functions 
(see \cite{Levin87}, \cite{Levin}, \cite{Goldreich}, and 
\cite{Trevisan}). 

The function ${\sf ev}_{\sf circ}$ is {\it polynomially balanced}. 
Indeed, for any input $X = ({\sf code}(C), x)$ and output 
$Y = ({\sf code}(C), C(x))$ of ${\sf ev}_{\sf circ}$ we have: 
 \ $|X| = |{\sf code}(C)| + |x| \ \leq \ 2 \ (|{\sf code}(C)| + |C(x)|)$
$ = 2 \, |Y|$, 
and \ $|Y| = |{\sf code}(C)| + |C(x)| \ \leq \ $
$2 \ (|{\sf code}(C)| + |x|) = 2 \, |X|$, 
using the facts that \ $|x| \leq |C|$, \ $|C(x)| \leq |C|$, and 
 \ $|C| \leq |{\sf code}(C)|$. 
Also, if $c$ is not the code of any circuit then
${\sf ev}_{\sf circ}(c,x) = (c,x)$, so length is preserved in that
case.

The function ${\sf ev}_{\sf circ}$ is not length-equality preserving,
therefore we introduce a special evaluation function
${\sf ev}_o: \{0,1\}^* \times \{0,1\}^* \longrightarrow $
$\{0,1\}^* \times \{0,1\}^*$, 

\medskip

${\sf ev}_o(c,x) \ = \ \left\{  \begin{array}{ll} 
  (c, \ C(x)) \ \ \ \ &  
        {\rm if} \ \ c = {\sf code}(C), 
        \ \ |c| \leq 12 \, m_C \, \log_2 (2 \, m_C), 
        \ {\rm and} \ \ |x| = m_C = n_C \, ,   \\  
   (c,x)               & {\rm otherwise}. 
            \end{array} \right. $

\medskip

\noindent This definition makes ${\sf ev}_o$ {\it length-preserving}, 
hence it is also length-equality preserving and polynomially balanced. 
Clearly, ${\sf ev}_o$ is also uniformly computable in polynomial time.
The definition was made in such a way that Prop.\ \ref{ClinsizeinN} 
can be applied.

\begin{lem} \label{evOneway} \  If 
 \, $\Pi_2^{\sf P} \neq \Sigma_2^{\sf P}$ then the special 
evaluation function ${\sf ev}_o$ is one-way by circuit size.
\end{lem}
{\bf Proof.} \ By contraposition, let us assume that ${\sf ev}_o$ has an 
inverse function ${\sf ev}'_o$ which is computed by a polynomial-size
family of circuits ${\bf E}' = (E'_i : i \in \varmathbb{N})$.
So, there is a polynomial $p(.)$ such that for all $i$, 
 \ $|E'_i| \leq p(i)$.  
The circuit $E'_i$ takes inputs of the form 
$(c,y) \in \{0,1\}^* \times \{0,1\}^*$ with $i = |c| + |y|$. 
Consider the case where $c = {\sf code}(C)$ for any circuit $C$ such 
that $m_C = n_C = |y|$, and $|c| \leq 12 \, m_C \, \log_2 (2m_C)$. 
Then \ $i = |c| + n_C = |c| + m_C$.
We let \ $C' = E'_i({\sf code}(C), \ \cdot)$ ; \, this is the circuit 
$E'_i$ with the $c$-input hardwired to the value ${\sf code}(C)$.  
Then the existence of an inverse $C'$ for every circuit $C$ as in 
Prop.\ \ref{ClinsizeinN}, implies $\Pi_2^{\sf P} = \Sigma_2^{\sf P}$.
 \ \ \ \ \ $\Box$

\medskip

\noindent Lemma \ref{evOneway} immediately implies: 

\begin{thm} \label{existsOneWayFunc} \ 
If $\Pi_2^{\sf P} \neq \Sigma_2^{\sf P}$ then there exist 
length-preserving functions that are {\em one-way by circuit size} 
and computable uniformly in polynomial time. 
 \ \ \ \ $\Box$ 
\end{thm}


\bigskip

\bigskip



\bigskip

\bigskip

\noindent {\bf Jean-Camille Birget} \\
Dept.\ of Computer Science \\
Rutgers University at Camden \\
Camden, NJ 08102, USA \\
{\tt birget@camden.rutgers.edu}

\end{document}